\documentclass{article}
\usepackage{latexsym}
\usepackage{amssymb}
\usepackage{amsfonts}
\usepackage{amsmath}
\usepackage[dvips]{graphicx}

\begin{document}

\title{Fitting the luminosity data from type Ia supernovae in the frame of
the Cosmic Defect theory}
\author{A. Tartaglia$^{1}$, M. Capone$^{1}$, V. Cardone$^{2}$, N. Radicella$%
^{1}$ \\
$^{1}$Dipartimento di Fisica del Politecnico, and INFN, Torino. \\
$^{2}$Dipartimento di Fisica dell'Universit\`{a} di Salerno, and INFN,
Salerno.}
\maketitle

\begin{abstract}
The Cosmic Defect (CD) theory is reviewed and used to fit the data for the
accelerated expansion of the universe, obtained from the apparent luminosity
of 192 SnIa's. The fit from CD is compared with the one obtained by means of 
$\Lambda $CDM. The results from both theories are in good agreement and the
fits are satisfactory. The correspondence between both approaches is
discussed and interpreted.
\end{abstract}

\section{\protect\bigskip Introduction}

As it is well known, an extremely important finding of the last decade has
been the accelerated expansion of the universe. This was rather a surprise,
mainly based on the observation of luminosity distance of type Ia supernovae
(SnIa) \cite{scoperta}. Nowadays, the picture which seems to emerge from the
data is that of an universe which has undergone a transition from a
decelerated to an accelerated phase, with a relatively recent turning point
located at $z_{tr}\simeq 0.46$ \cite{dec vs acc}. This framework seems to be
confirmed by cross-comparison with other pieces of evidence \cite{cmb}. The
discovery gave start to an active search for an explanation on the
theoretical side, within and outside general relativity (GR). An immediate
effect was to revive the old cosmological constant, $\Lambda $ \cite%
{costante}; afterwards, a number of evolutionary sons of $\Lambda $ or new
exotic fields were elaborated, mostly based on the idea of "dark energy" 
\cite{dark}\cite{phantom}\cite{K-essence}\cite{tachyon}\cite{chaplygin}.
Also various possibilities of alternative, modified or extended versions of
GR have actively been explored \cite{ET and AT}.

Here, our purpose is to review the existing observational data and some
proposed fits, comparing them one with another and with the results of a
recently introduced four-vector theory, that we shall call the "cosmic
defect" theory, CD for short \cite{nostro}. The CD theory, which has also
correspondences in the group of the so called vector "ether" theories \cite%
{vector}, will also be revised and recast in the following.

Whenever a theory is contrasted with the data from experience (here, from
observation) one has to face a number of different problems. First of all,
comes the reliability and cleanness of the data: we shall not elaborate on
this, assuming the discussion to have been effectively conducted in the
literature \cite{errori}. A second subtle issue is that, even in presenting
apparently raw data, underlying assumptions often exist, originating in one
or another theoretical view: as far as possible, we shall try to express the
existing information in a model-independent way. Finally any theory usually
has (a number of) free parameters to adjust, in order to fit the experiment;
of course the more parameters you have, the more you will be able to
reproduce a given empirical trend, however any choice must be checked for
consistency in as many different physical situations as possible.

As we shall see, the CD theory gives a reasonably good fit for the SnIa
data, making use of a limited number of parameters and, in the same time,
offers an interpretation paradigm based on correspondences with known
physical phenomena without the need of calling in new dark entities.

\section{Luminosity distance, magnitude and redshift}

In the framework of the supernovae observations, a key role is played by the
concept of luminosity distance, $d_{l}$, which is defined as 
\begin{equation}
d_{l}\doteq \sqrt{\frac{L_{obs}}{4\pi \Phi }}.  \label{dl}
\end{equation}%
$L_{obs}$ is the absolute luminosity of the source (released energy per unit
time) corresponding to the $z$ value measured by the observer, $\Phi $ is
the energy flux density (energy per unit time and surface) measured at the
observer's site. In an expanding universe both energy and time are affected
by the expansion so that the effective luminosity for the observer, in terms
of the absolute luminosity at the source is \cite{luminosity}:%
\[
L_{obs}=\frac{L_{S}}{\left( 1+z\right) ^{2}}. 
\]%
In an universe endowed with the typical Robertson Walker (RW) symmetries (%
\ref{dl}) becomes 
\[
d_{l}=a_{0}r_{S}(1+z). 
\]%
where $a_{0}$ is the scale parameter at the observer, and $r_{S}$ is the
coordinate distance of the source from the observer. The latter, written in
terms of the distance travelled by a light ray, is in turn%
\[
r_{S}=c\int_{t_{S}}^{t_{0}}\frac{dt}{a(t)}, 
\]%
where of course $t$ is the cosmic time. In terms of the redshift and the
scale factor we may also write 
\begin{equation}
cdt=c\frac{da}{\dot{a}}=-\frac{cdz}{\left( 1+z\right) H(z)},
\label{dt func dz}
\end{equation}%
where the dot denotes the derivative with respect to $t$ and $H=\dot{a}/a$
is the Hubble parameter.

It is then easily seen that the luminosity distance is%
\[
d_{l}=c(1+z)\int_{0}^{z}\left( 1+\zeta \right) \frac{da\left( \zeta \right) 
}{\dot{a}\left( \zeta \right) }=c(1+z)\int_{0}^{z}\frac{d\zeta }{H(\zeta )}. 
\]

Usually astronomical objects are classified in terms of their magnitude $m$,
rather than their luminosity. By definition, the bolometric magnitude
(integrated over all frequencies) depends logarithmically on the luminosity
distance, according to the formula: 
\begin{eqnarray}
m-M_{S} &=&25+5\log d_{l}=25+5\log \left( a_{0}c\left( 1+z\right)
\int_{t_{S}}^{t_{0}}\frac{dt}{a}\right)  \nonumber \\
&=&25+5\log \left( c\left( 1+z\right) \int_{0}^{z}\frac{d\zeta }{H(\zeta )}%
\right)  \label{magnitudo1} \\
&=&25+5\log \left( c\frac{\left( 1+z\right) }{H_{0}}\int_{0}^{z}\frac{d\zeta 
}{E(\zeta )}\right)  \nonumber
\end{eqnarray}%
where distances are expressed in Mpc and it is $H_{0}=H\left( 0\right) $ and 
$E\left( z\right) =H\left( z\right) /H_{0}$; $m-M_{S}$ is usually called the
"distance modulus".

The integral in (\ref{magnitudo1}) depends of course on the model one uses
to describe the cosmic expansion. For a dust filled universe in a typical
Friedman-Robertson-Walker (FRW) scenario it is indeed 
\begin{equation}
a\left( t\right) =a_{0}\sqrt[3]{6\pi G\rho _{m0}}t^{2/3},  \label{friedman}
\end{equation}%
being $\rho _{m0}$ the present matter energy density and $G$ the gravitation
constant.

As a consequence one expects%
\begin{equation}
\left( m-M_{S}\right) _{FRW}=25+5\log \left[ \frac{3c}{\sqrt{6\pi G\rho _{m0}%
}}\left( 1+z-\sqrt{1+z}\right) \right]  \label{FRW}
\end{equation}

If one considers a $\Lambda $-cold-dark-matter universe ($\Lambda CDM$),
i.e. an FRW universe with a cosmological constant $\Lambda $, it is%
\begin{equation}
E\left( z\right) =\sqrt{\Omega _{m}\left( 1+z\right) ^{3}+1-\Omega _{m}}
\label{LCDM}
\end{equation}%
where $\Omega _{m}=\rho _{m}/\rho _{c}$ represents the ratio between the
matter density and the critical density (ensuring the flatness of space).
The difference $\Omega _{\Lambda }=1-\Omega _{m}$ allows for the effect of
the cosmological constant.

Formula (\ref{LCDM}) is a special case of the more general 
\[
E\left( z\right) =\sqrt{\sum_{i}\Omega _{i}\left( 1+z\right) ^{3(1+w_{i})}}, 
\]%
allowing for any number of components of the content of the universe, with
different equations of state.

\section{\protect\bigskip The Cosmic Defect theory}

The CD theory is based on the presence of a cosmic (four)-vector field in
the universe. This vector field is interpreted as the strain flux density in
a continuum with a pointlike defect\footnote{%
Actually the defect could correspond to any singular hypersurface.} \cite%
{nostro}. The pointlike nature of the defect induces the RW symmetry; this
very symmetry, together with the defect paradigm, implies the vector to be
"radial", i.e. everywhere parallel to the cosmic time axis, and
divergence-free, except at the defect. The norm of the vector, $\chi $, will
coincide with the absolute value of its time component and, according to the
divergence-free feature, will be:%
\begin{equation}
\chi =\frac{Q^{3}}{a^{3}},  \label{chi}
\end{equation}%
where $Q$ is a constant and $a$ is the scale factor of the RW metric.

The other relevant feature of the CD\ theory is in the choice of the
Lagrangian for the spacetime containing the defect. This choice is inspired
by the correspondence between the (bidimensional) phase-space of a RW
universe and the one of a point particle moving through a viscous fluid \cite%
{nostro}. Including the presence of "matter" (i.e. whatever is not accounted
for by spacetime), the action integral is%
\begin{equation}
S=\int \left( \kappa e^{-g_{\mu \nu }\gamma ^{\mu }\gamma ^{\nu }}R+\mathcal{%
L}_{matter}\right) \sqrt{-g}d^{4}x,  \label{azione}
\end{equation}%
with $\kappa \equiv c^{4}/16\pi G$ and $d^{4}x=dtdrd\theta d\varphi .$
Explicitly introducing the RW symmetry and considering matter in terms of
scalar functions, the Lagrangian read out of (\ref{azione}) is:%
\begin{equation}
\mathcal{L}_{0}\mathcal{=}-\mathcal{V}_{k}\left[ 6\kappa e^{-\chi
^{2}}\left( a^{2}\ddot{a}+a\dot{a}^{2}\right) +\kappa _{0}fa^{3}\dot{a}%
^{2}+\varpi ha^{3}\right] ,  \label{lagra0}
\end{equation}%
where $\mathcal{V}_{k}$ is the part of the Lagrangian which is not affected
by any variation with respect to the metric, and, in the flat $k=0$ case
(polar coordinates), equals $r^{2}\sin \theta $. The presence of matter is
represented by two scalar functions, $f$ and $h$, coupling to spacetime
through the constants $\kappa _{0}$ and $\varpi $. The $f$ function accounts
for a possible coupling to the rate of expansion of the universe, $\dot{a}$.

The second derivative of $a$ with respect to $t$, appearing in (\ref{lagra0}%
), is easily eliminated, once the action is integrated by parts, thus giving
a final effective Lagrangian of the universe

\begin{equation}
\mathcal{L}=-\mathcal{V}_{k}\left[ -6\kappa e^{-\chi ^{2}}\left( \frac{6}{%
a^{5}}+a\right) \dot{a}^{2}+\kappa _{0}fa^{3}\dot{a}^{2}+\varpi ha^{3}\right]
.  \label{lagreff}
\end{equation}

From (\ref{lagreff}) the Hamiltonian function is readily obtained,

\begin{equation}
\mathcal{H}\doteq \dot{a}\frac{\partial \mathcal{L}}{\partial \dot{a}}-%
\mathcal{L}=-\mathcal{V}_{k}\left\{ \left[ \kappa _{0}fa^{3}-6\kappa
e^{-\chi ^{2}}\left( \frac{6}{a^{5}}+a\right) \right] \dot{a}^{2}-\varpi
ha^{3}\right\} .  \label{hamil}
\end{equation}%
As usual, $\mathcal{H}$ can be interpreted as the energy content of the
system described by the effective Lagrangian (\ref{lagreff}), so that in our
case it represents the energy content of the universe. The Hamiltonian of an
isolated system is a conserved quantity, since it is identically 
\begin{equation}
\frac{d\mathcal{H}}{dt}=\ddot{a}\frac{\partial \mathcal{L}}{\partial \dot{a}}%
+\dot{a}\frac{d}{dt}\frac{\partial \mathcal{L}}{\partial \dot{a}}-\frac{%
\partial \mathcal{L}}{\partial a}\dot{a}-\frac{\partial \mathcal{L}}{%
\partial \dot{a}}\ddot{a}\equiv 0  \label{conservazione}
\end{equation}%
From now on use will be made of $\alpha =a/Q$ so we write 
\begin{equation} \label{costante}
\left[ \kappa _{0}f\alpha ^{3}-6\kappa e^{-\chi ^{2}}\left( \frac{6}{\alpha
^{5}}+\alpha \right) \right] \dot{\alpha}^{2}-\varpi h\alpha ^{3}\mathcal{=W=%
}\textit{ const.} 
\end{equation}

From (\ref{costante}) one directly gets the expansion rate equation:%
\[
\dot{\alpha}^{2}=\frac{\mathcal{W}+\varpi h\alpha ^{3}}{\kappa _{0}f\alpha
^{3}-6\kappa e^{-\chi ^{2}}\left( \frac{6}{\alpha ^{5}}+\alpha \right) }. 
\]

Actually, if we want to recover the usual meaning of the matter term in a
comoving reference frame, we must choose%
\[
\kappa _{0}=0, 
\]%
so that the expansion rate can be rewritten as%
\begin{equation}
\dot{\alpha}^{2}=-\frac{\mathcal{W}+\varpi h\alpha ^{3}}{6\kappa
e^{-1/\alpha ^{6}}\left( \frac{6}{\alpha ^{5}}+\alpha \right) }.
\label{field}
\end{equation}%
In the absence of a defect we should recover the classical FRW model; for
this reason, it should be%
\begin{equation}
\frac{\dot{a}^{2}}{a^{2}}=\frac{8\pi G}{3}\rho =\frac{1}{6\kappa }\rho c^{4},
\label{uno}
\end{equation}%
where $\rho c^{2}$ is the energy density of matter. However, under the same
condition ($Q=\chi =0$) eq. (\ref{field}) gives%
\begin{equation}
\frac{\dot{a}^{2}}{a^{2}}=-\frac{\mathcal{W}Q^{3}+\varpi ha^{3}}{\kappa a^{3}%
}=-\frac{\varpi }{\kappa }h.  \label{due}
\end{equation}%
Consistency between (\ref{uno}) and (\ref{due}) then requires%
\begin{eqnarray*}
\varpi &=&-1/6, \\
h &=&\rho c^{4}.
\end{eqnarray*}

The final formula for the expansion rate of the universe is

\begin{equation}
\dot{\alpha}^{2}=-\frac{\mathcal{W}-\rho c^{4}\alpha ^{3}}{6\kappa
e^{-1/\alpha ^{6}}\left( \frac{6}{\alpha ^{5}}+\alpha \right) }.
\label{adot}
\end{equation}%
Let us now suppose that the cosmic fluid is made of a number of different
non-interacting components, each with its equation of state in the form%
\[
p_{i}=w_{i}\rho _{i}c^{2}, 
\]%
where $w_{i}$ are real positive numbers $(w_{i}\geq 0)$, and $p_{i}$ is the
partial pressure of the $i_{th}$ component.

The conservation laws imply that%
\[
\rho _{i}=\rho _{i0}\frac{\alpha _{0}^{3\left( 1+w_{i}\right) }}{\alpha
^{3\left( 1+w_{i}\right) }}. 
\]%
Introducing this relation into (\ref{adot}) we have: 
\begin{equation}
\dot{\alpha}^{2}=-\frac{\mathcal{W}-c^4\sum\limits_{i}\rho _{i0}\frac{
\alpha _{0}^{3\left( 1+w_{i}\right) }}{\alpha^{3w_{i}}}}{6\kappa
e^{-1/\alpha ^{6}}\left(\frac{6}{\alpha ^{5}}+\alpha \right) },
\label{dustpiurad}
\end{equation}%
The corresponding Hubble parameter is:

\begin{eqnarray}
H &=&\frac{\dot{a}}{a}=\frac{\dot{\alpha}}{\alpha }=\frac{1}{\alpha }\sqrt{%
\frac{c^{4}\sum\limits_{i}\rho _{i0}\frac{\alpha _{0}^{3\left(
1+w_{i}\right) }}{\alpha ^{3w_{i}}}-\mathcal{W}}{6\kappa e^{-1/\alpha
^{6}}\left( \frac{6}{\alpha ^{5}}+\alpha \right) }}  \nonumber \\
&=&\frac{c^{2}}{\sqrt{6\kappa }}\left( 1+z\right) ^{3/2}\sqrt{\frac{%
\sum\limits_{i}\rho _{i0}\left( 1+z\right) ^{3w_{i}}-\mathfrak{w}}{%
e^{-\left( 1+z\right) ^{6}/\alpha _{0}^{6}}\left[ 1+6\left( 1+z\right)
^{6}/\alpha _{0}^{6}\right] }}  \label{da integrare}
\end{eqnarray}%
with $\mathfrak{w=}\frac{\mathcal{W}}{c^{4}\alpha _{0}^{3}}$.

In the case of dust ($w=0$) and radiation ($w=1/3$) it is%
\begin{equation}
H\left( z\right) =\left( 1+z\right) ^{3/2}\sqrt{\frac{c^{4}}{6\kappa }\rho
_{m0}}\sqrt{\frac{1+\varepsilon _{0}\left( 1+z\right) -b}{e^{-\left(
1+z\right) ^{6}/\alpha _{0}^{6}}\left[ 1+6\left( 1+z\right) ^{6}/\alpha
_{0}^{6}\right] }}  \label{perilfit}
\end{equation}%
The adimensional quantity $\varepsilon _{0}=\rho _{r0}/\rho _{m0}$ is the
present ratio between the radiation and the matter energy density in the
universe. $b$ is $\mathfrak{w/}\rho _{m0}$.

\section{Observations vs theory}

In order to compare theory and observation we make reference to the same set
of data used recently by Davis et al. \cite{davies} and incorporating
supernovae analyzed in four different groups: 60 from the ESSENCE (Equation
of State: SupErNovae trace Cosmic Expansion) project \cite{wood}, 57 from
SNLS (SuperNova Legacy Survey) \cite{astier}, 45 nearby supernovae, 30
detected by the Hubble Space Telescope and qualified as "golden" supernovae
by Riess et al. \cite{riess}. As mentioned in the introduction, we shall not
enter into the discussion of the elaboration of the data, but assume them
exactly the way they are published or anyway accessible considering them as
the best available at the moment.

Altogether we use the luminosity data from 192 SnIa \cite{dati} which we try
to fit with theoretical models.

\begin{figure}[tbp]   
 \begin{center}
   \includegraphics[width=13 cm]{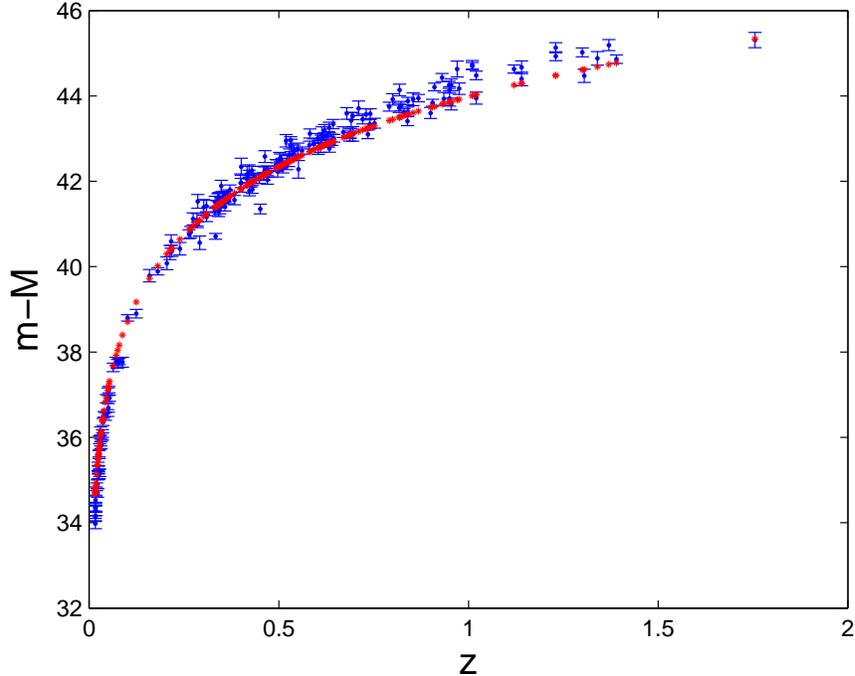}
\caption{ \label{FRWgraph}Fit of distance modulus observations using a 
   standard dust filled Friedman-Robertson-Walker universe. The data are 
   from 192 SnIa's as explained in the text. Vertical bars represent the 
   experimental uncertainties ( $2$ $\sigma$). The uncertainty on the 
   redshift parameter $z$ would be unperceivable at the scale of the graph. 
   The reduced $\chi ^{2}$ is 2.1276. }
 \end{center}
\end{figure}

First we use (\ref{FRW}) and obtain the result shown on fig. \ref{FRWgraph}.%

The direct inspection of the graph shows that the data correspond to
systematically lower luminosities than the ones given by the FRW model,
whence the accelerated expansion interpretation comes.

The next step will be to test on the data the $\Lambda CDM$ model in its
simplest version. To that purpose we use (\ref{magnitudo1}) and (\ref{LCDM}%
). In practice%
\begin{equation}
m-M_{S}=\mu +5\log \left( 1+z\right) +5\log \int_{0}^{z}\frac{d\zeta }{\sqrt{%
\Omega _{m}\left( 1+\zeta \right) ^{3}+1-\Omega _{m}}}  \label{LCDM1}
\end{equation}%
The result, as it is well known, is better than before, since the reduced $%
\chi ^{2}$ now is $\chi ^{2}=1.0295$ with a best fitting $\mu =43.30\pm 0.03$%
, which correspons to $H_{0}=65.6\pm 0.9$ km/s$\times $Mpc, and $\Omega
_{m}=0.27\pm 0.03$, i.e. 27\% of ordinary and dark matter plus 73\% of dark
energy (cosmological constant) in a spatially flat universe. For the
optimization as well as for the determination of the uncertainty of the
values of the parameters, use has been made, as for the previous FRW case,
of a multidimensional nonlinear minimization by means of the MINUIT engine 
\cite{minuit}\footnote{%
The open source routine we used, due to G. Allodi of the university of
Parma, is named fminuit, is called from within MATLAB, and may be retrieved
from ftp://ftp.fis.unipr.it/pub/matlab/fminuit.mex}. The optimization is
made minimizing the reduced $\chi ^{2}$ of the fit.

\begin{figure}[t]   
 \begin{center}
   \includegraphics[width=13 cm]{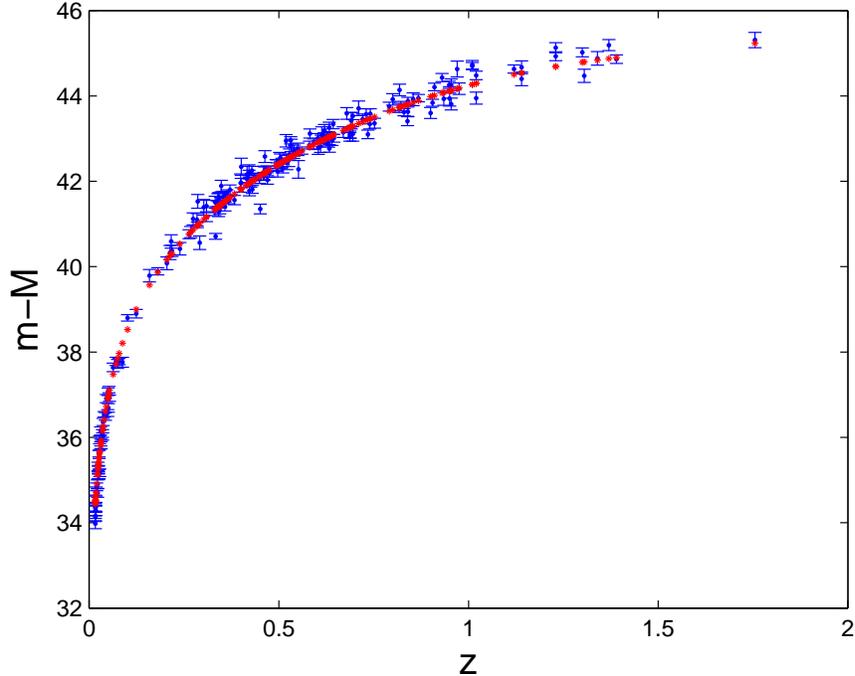}
   \caption{\label{CD4fit} Fit of distance modulus
     observations using the CD theory with two free parameters. Radiation is
     overlooked with respect to dust. Symbols as in Fig.~\ref{FRWgraph}. The 
     reduced $\chi^2$ is 1.092.}
 \end{center}
\end{figure}

Finally we test the CD theory. Use is made of (\ref{magnitudo1}) and (\ref%
{perilfit}) considering dust and radiation, so that

\begin{equation}
m-M_{S}=\mu +5\log \left( 1+z\right) +5\log \int_{0}^{z}\sqrt{\frac{%
e^{-\left( 1+\zeta \right) ^{6}/\alpha _{0}^{6}}\left[ 1+6\left( 1+\zeta
\right) ^{6}/\alpha _{0}^{6}\right] }{\left( 1+\zeta \right) ^{3}\left(
1+\varepsilon _{0}\left( 1+\zeta \right) -b\right) }}d\zeta  \label{CD}
\end{equation}%
We could treat $\mu $, $\alpha _{0}$ and $b$ as optimization parameters,
however the value to be introduced for $\varepsilon _{0}$ is the one
currently agreed upon, excluding any dark contribution: $\varepsilon
_{0}\sim 10^{-4}$. Of course as far as $z$ is in the order of a few units
(as it is the case for SnIa's) the radiation term in the denominator of the
integrand is negligible, so that also the contribution of $b$ may be
embedded in $\mu $ and the free parameters remain $\mu $ and $\alpha _{0}$
only. The result of the optimization process is $\mu =43.26\pm 0.03$ and $%
\alpha _{0}=1.79\pm 0.04$; the reduced $\chi ^{2}$ is $\chi ^{2}=1.092$,
almost as good as for $\Lambda CDM$. The graph is shown in fig. \ref{CD4fit}.

\subsection{Effective gravitational coupling constant}

The meaning of the choice of SnIa's for cosmological analyses relies on the
assumption that they are "standard candles", which means that the relation
between their light curve and luminosity is thought to be always the same,
now as well as at epochs corresponding to high cosmic redshifts (actually,
at the moment, the highest redshift for an SnIa corresponds to $z\simeq 1.8$%
). If it is so, in an expanding universe we expect any standard candle (as
an SnIa is assumed to be) to appear dimmer and redder according to the
peculiar expansion law.

Indeed the currently assumed explosion mechanism for Ia-type supernovae
makes their peak luminosity be proportional to the mass of nickel
synthesized, which, to a good approximation, is a fixed fraction of the
Chandrasekhar mass \cite{nickel}. The Chandrasekhar mass, in its turn, is
proportional to $G^{-3/2}$ \cite{varG}. In an expanding universe it is
reasonable to expect that an effective value of $G$, somehow accounting for
the expansion, must replace the usual constant in the formulae we use to
describe locally the behaviour of matter under gravity. If it is so a
consequence is that the absolute magnitude of a source at a given $z$
differs from the one for $z=0$. In practice, including the fact that the
extension of the light curve (characteristic time) of an SnIa is
approximately proportional to the square root of the Chandrasekhar limit 
\cite{trequarti}, it is 
\[
M\left( 0\right) _{S}-M_{S}\left( z\right) =\frac{15}{4}\log \frac{G\left(
z\right) }{G_{0}}, 
\]%
and the distance modulus becomes:

\[
m(z)-M_{S}=25+5\log d_{l}+\frac{15}{4}\log \frac{G\left( z\right) }{G_{0}}. 
\]

For the above reason it is important, in any model we use in order to
describe the evolution of the universe, to determine the explicit form of $%
G\left( z\right) $. This is indeed a delicate task, since everything has to
be worked out in terms of appropriate approximations. Attempts have been
made to solve this problem in general terms for any scalar-tensor theory
leading to acceptable results for limited ranges of $z$ values \cite%
{nesseris}. We would like to follow a different path.

Actually at the scale of a stellar system the effect of the cosmic curvature
appears only as a very tiny perturbation of the usual stationary state
solutions of general relativity. Considering for instance the example of a
spherically distributed bunch of matter in a background RW universe, we
expect that locally the induced metric be essentially 
\begin{equation}
ds^{2}=\left( 1-f\left( r,\mathfrak{t}\right) \right) d\mathfrak{t}%
^{2}-a^{2}\left( \mathfrak{t}\right) \left( \frac{dr^{2}}{1-h\left( r,%
\mathfrak{t}\right) }+r^{2}d\theta ^{2}+r^{2}\sin ^{2}\theta d\phi
^{2}\right)  \label{lineanewton}
\end{equation}%
with an extremely weak dependence of $f$ and $h$ on time also. If $\tau _{0}$
is the given cosmic time we use $\mathfrak{t}=\tau -\tau _{0}$ as the time
variable now. Far away from the local source it must be $f$, $h\rightarrow 0$
so that the pure RW metric is recovered. On the other side if it were $%
a=a_{0}=$ constant we could incorporate its value in the definition of $r$
and from the field equations we would obtain the Schwarzschild solution $%
f=h=2Gm/c^{2}r$.

As written above, $a_{0}$ may be absorbed into a rescaling of $r$ and
similarly $G_{0}$ may be included into $f$ and $h$. Now we may think to
develop $a,$ $f,$ and $h$ in a power series of $\mathfrak{t}$; if we do so,
including the condition $f$, $h<<1$, we end up \cite{ATNR} with the
interesting result that the zero order (in the typical evolution times of
the gravitational phenomena we wish to describe) is independent from such
parameters as Hubble's. In practice rapid events, such as the accretion of
matter preluding to a gravitational collapse, always happen to be controlled
by the universal value of $G$ and no correction is expected on the distance
modulus. Any effect of the expansion of the universe (differentiating one
model from another) emerges only for times comparable to the cosmic
evolution times.

\section{The Hubble parameter and the age of the universe}

Reconsidering now the explicit spelling of the parameters appearing in the
CD theory used to draw fig. \ref{CD4fit} we see that it is%
\[
\mu =25-5\log c+\frac{5}{2}\log \frac{6\kappa }{\rho _{m0}\left( 1-b\right) }
\]%
whence%
\begin{eqnarray}
\rho _{m0}\left( 1-b\right) &=&\frac{6\kappa }{c^{2}}10^{-\frac{2}{5}\left(
\mu -25\right) }  \label{rom0} \\
&=&\left( 8.\,\allowbreak 5\mp 0.2\right) \times 10^{-27}\text{ kg/m}^{3} 
\nonumber
\end{eqnarray}%
The "visible" matter density in the universe is commonly assumed to be
around $\sim 10^{-27}\div 10^{-28}$ kg/m$^{3}$ which means that $b$ must be $%
\sim -10$.

Then, introducing (\ref{rom0}) into (\ref{perilfit}) and evaluating for $z=0$
we obtain 
\begin{eqnarray*}
H_{0} &=&\sqrt{\frac{c^{4}}{6\kappa }\frac{\rho _{m0}\left( 1+\varepsilon
_{0}-b\right) }{e^{-1/\alpha _{0}^{6}}\left( 1+6/\alpha _{0}^{6}\right) }} \\
&=&\left( 62.\,\allowbreak 8\mp 1.7\right) \text{ km/}\left( \text{s}\times 
\text{Mpc}\right)
\end{eqnarray*}%
which is an acceptable result ($\varepsilon _{0}$ has been neglected with
respect to $b$). The corresponding Hubble time is 15.6 Gy.

Of course we should determine the age of the universe using the CD model;
this can be done by means of (\ref{dustpiurad}) by integration:

\begin{eqnarray*}
t_{0} &=&\frac{1}{c^{2}}\sqrt{\frac{6\kappa }{\rho _{m0}\alpha _{0}^{3}}}%
\int_{0}^{\alpha _{0}}\sqrt{\frac{\left( 6+\xi ^{6}\right) e^{-\frac{1}{\xi
^{6}}}}{\xi ^{4}\left[ \left( 1-b\right) \xi +\varepsilon _{0}\alpha _{0}%
\right] }}d\xi \\
&=&\allowbreak \left( 9.\,\allowbreak 0\pm 0.2\right) \text{ Gy}
\end{eqnarray*}%
The final numerical result has been obtained neglecting $\varepsilon
_{0}\alpha _{0}$ with respect to the other terms in the denominator of the
integrand. The value falls rather short as compared to the age of globular
clusters, which fact may probably be interpreted as an inadequacy of the
model at very early cosmic times.

\section{Conclusion and discussion}

We have fitted the apparent luminosity data from SnIa's with the values
predicted by the $\Lambda $CDM and the CD theories, comparing both with a
traditional FRW universe. The result is of course partly known, but we see
now that also CD improves with respect to FRW and gives a fit comparable
with the one of $\Lambda $CDM. Using the same data and the same number of
parameters we obtained similar values of the riduced $\chi ^{2}$'s
suggesting the idea that CD also is a viable theory. It is however true that
the apparently small difference of the reduced $\chi ^{2}$'s of the fits
corresponds to a rather big difference in the full $\chi ^{2}$ that, when
analyzed on the light of statistical information criteria, such as the
Akaike Information Criterion (AIC) \cite{aic} and the Bayesian Information
Criterion (BIC) \cite{bic}, enhances the distance between the two theories
in favor of $\Lambda $CDM. In the same time it is also true that both
reduced values of $\chi ^{2}$ are bigger than $1$; furthermore the $H_{0}$
values obtained from observation using different methods are systematically
higher than the ones of the $2$-parameters best fits above. The most recent
data from WMAP\ \cite{wmap} yield $H_{0}=73.2_{-3.2}^{+3.1}$ km/s$\times $%
Mpc which is consistent with a number of other results produced by different
methods and indicators (like SnI, SnII, Cepheids in nearby galaxies,
Sunyaev-Zeldovitch effect, X rays from clusters, gravitationally lensed
systems) all quoted in \cite{wmap}. The central values from these different
observations range from $72$ to $76$ km/s$\times $Mpc and in general the
historical evolution of the estimated values of the Hubble constant seems to
progressively converge towards something around $75$ km/s$\times $Mpc%
\footnote{%
Look for instance at http://cfa-www.harvard.edu/\symbol{126}huchra/hubble/},
which is $\sim 15\%$ more than the results got by means of the fits in this
paper. If the "experimental" value of $H_{0}$ were used in the fits (so
reduced to $1$-parameter ones) the agreement with the data would
consistently worsen both for $\Lambda $CDM and for CD. In practice there is
something missing beyond the details of the theories and their
interpretation, which deserves investigation and insight.

The $\Lambda $CDM is indeed different from the CD theory: the former assumes
in the universe the presence of a cosmological constant corresponding to a
sort of uniformly and homogeneously distributed dark energy;\ the latter
interprets space-time as a continuum with a cosmic defect inducing a
strained state containing both the symmetry and the non-uniform expansion
rate. Besides this, we know that $\Lambda $CDM requires also that the matter
content in the universe be one order of magnitude bigger than what expected
from baryonic particles only. In the case of the CD theory, instead, we saw
that the ordinary matter density is combined with the effect induced by the
defect via the $b$ parameter (see \ref{rom0}), so that, in a sense, it gives
rise to an effective matter/energy density one order of magnitude bigger
than the actual one. Adding the fact that one can interpret the strained
state induced by the cosmic defect as being the equivalent of a non-uniform
(in time) dark energy, we see that in facts the principle difference between 
$\Lambda $CDM and CD could not be that deep. The CD theory already proved to
correspond to vector theories developed with different motivations and
within a different scenario \cite{vector}. The real difference between the
theories, not considering the details, is in the end in the fact that CD
tries to give an "explanation" for the dark energy, that seems to permeate
the universe, within a consistent paradigm, which is the one of the defected
continuum with its properties.

Of course this test of the CD theory is limited to the SnIa data and
consequently to a limited range of $z$ values, and the poor result obtained
for the age of the universe seems to indicate an inadequacy of the theory at
high redshift values, where probably a better treatment of the matter
content is in order. The result with the type Ia supernovae is however
encouraging.

\end{document}